\documentstyle[aps,floats,epsf,epsfig]{revtex}
\begin{document}

\font\el=cmbx10 scaled \magstep2 {\obeylines \hfill NCKU-HEP-99-05
\hfill \hfill{hep-ph/9904449}}

\vskip 1.5 cm


\begin{center}
{\large {\bf Perturbative QCD analysis of $b$-hadron lifetimes}}
\vskip 0.8cm
Hsiang-nan Li
\vskip 0.1cm
Department of Physics, National Cheng-Kung University,\par
Tainan, Taiwan 701, Republic of China
\vskip 0.3cm
\end{center}
\vskip 0.5cm


\vskip 0.5cm
\centerline{\bf Abstract}
\vskip 0.3cm

We develop perturbative QCD factorization theorems for inclusive $b$-hadron 
decays, in which radiative corrections characterized by the hadronic
scale, the $b$-hadron mass, and the $W$ boson mass are absorbed into a
heavy hadron distribution function, a hard $b$ quark decay amplitude, and
a ``harder" function, respectively. Double logarithmic corrections
associated with a light energetic final-state quark, which appear at
kinematic end points, are absorbed into a jet function. Various large
logarithms contained in the above functions are summed to all orders,
leading to the evolution factors among the three characteristic scales.
The heavy hadron distribution function is identical to the one constructed
in the framework of heavy quark effective theory. It is shown that hadron
kinematics must be employed in factorization theorems, and that 
perturbative contributions, depending on hadron kinematics, distinguish the
lifetimes of the $b$-hadrons $B_d$, $B_s$ and $\Lambda_b$. Assuming the same
heavy-quark-effective-theory parametesr $\lambda_1$ for these hadrons, we
predict the lifetimes $\tau(B_d)=1.56$ ps, $\tau(B_s)=1.46$ ps and 
$\tau(\Lambda_b)=1.22$ ps. We also predict the $B_u$ meson lifetime 
$\tau(B_u)=1.62$ ps by varying the $B$ meson distribution function slightly. 
All the above results are consistent with experimental data.

\section{INTRODUCTION}

One of the puzzles in heavy hadron decays is the explanation of the low
lifetime ratio $\tau(\Lambda_b)/\tau(B_d)=0.79\pm 0.06$ \cite{ad}. 
Inclusive heavy hadron decays involve both nonperturbative and perturbative
corrections to the tree-level $b$ quark decay amplitudes. In the framework
of heavy quark effective theory (HQET) \cite{aa}, nonperturbative
corrections are expanded in powers of $1/M_b$, $M_b$ being the $b$ quark
mass, with the coefficient of each power proportional to a hadronic matrix
element of local operators, and perturbative corrections are evaluated order
by order at the quark level. The HQET prediction for the ratio
$\tau(\Lambda_b)/\tau(B_d)$ to $O(1/M_b^2)$ is about 0.99 \cite{NS}. When
including the $O(1/M_b^3)$ corrections, the ratio depends on six unknown
parameters, and reduces to around 0.97 for various model estimates, which
is still far beyond the experimental data. On the other hand, a
phenomenological ansatz was proposed, in which the overall $M_b^5$ factor
in front of nonleptonic decay widths is replaced by the corresponding
hadron mass $M_H^5$ \cite{ac}. This ansatz provides a solution to the
puzzle, since $(M_B/M_{\Lambda_b})^5=0.73$, $M_B$ and $M_{\Lambda_b}$ being
the $B$ meson mass and the $\Lambda_b$ baryon mass, respectively, is close
to the observed ratio. As pointed out in \cite{ae}, the same replacement
also explains the absolute $B$ meson decay rate, while the HQET prediction
using the expansion parameter $M_b$ accounts for only 80\% of the decay rate.

An alternative approach to inclusive heavy hadron decays is the perturbative
QCD (PQCD) factorization theorem \cite{ab}. In this formalism radiative
corrections to $b$ quark decays are absorbed into different factors in
factorization formulas for decay widths according to their characteristic
scales. The soft corrections characterized by the hadronic scale of order
$\Lambda_{\rm QCD}$ are factorized into a universal heavy hadron
distribution function. The corrections characterized by the heavy hadron
mass and by the $W$ boson mass are factorized into a hard $b$ quark decay
amplitude and a ``harder" function, respectively. The heavy hadron
distribution function provides a summation over the initial states of the
$b$ quark, such that the virtual and real soft gluon corrections 
cancel exactly. The introduction of a distribution function indicates that 
a residual momentum of the $b$ quark, whcih arises as an effect
from the light degrees of freedom in the heavy hadron, is allowed. The
inclusion of the light degrees of freedom then demands the use of the heavy
hadron kinematics, whose difference from the $b$ quark kinematics sets a
constraint on the magnitude of the residual momentum.

The soft gluons are collected into the distribution function
by employing an eikonal approximation, under which a $b$ quark
propagator is simplified into the propagator associated with the large
component of the rescaled $b$ quark field in HQET. This observation implies
the identity of the heavy hadron distribution function to the one
constructed in the HQET framework \cite{N}. Therefore, nonperturbative 
corrections also
start from $O(1/M_b^2)$ and the quark-hadron duality is respected in the
PQCD formalism, the same as in HQET. However, as transforming the $b$ quark
kinematics to the heavy hadron kinematics through $M_b=M_H-{\bar\Lambda}$,
the $O({\bar\Lambda}/M_H)$ correction occurs \cite{N}. It has been
explicitly demonstrated, taking the $B\to X_u l{\bar\nu}$ as an example
\cite{ab}, that the $O(1/M_B)$ correction from the $B$ meson distribution
function cancels that from the phase space factor $M_B^5$. 

As to perturbative corrections, various large logarithms contained in $b$
quark decays are summed to all orders in the PQCD approach. The results are
the renormalization-group (RG) evolutions among the three characteristic
scales, such that the factorization formulas are independent of the
renormalization scale $\mu$. Double logarithms associated with light
energetic final-state quarks are organized by the Collins-Soper resummation
technique \cite{CS}, leading to a Sudakov factor which smears end-point
singularities and improves the applicability of PQCD. These perturbative
factors, depending on kinematics of the initial heavy hadrons and of final
states, differ among various decay modes of a hadron and among various
hadrons. This is natural, because different decay modes should involve
different energy releases. In the HQET approach perturbative corrections are
evaluated to finite orders explicitly at the quark level with the 
renormalization scale $\mu$ set to a common value for all decay modes. It 
is then not a surprise that HQET predicts the lifetime ratio 
$\tau(\Lambda_b)/\tau(B_d)\approx 0.99$, since nonperturbative corrections, 
supposed to distinguish the decays of different $b$-hadrons, are suppressed 
by $1/M_b^2$, and perturbative corrections to $b$ quark decays are the same. 

The $B$ meson distribution function has been extracted from the photon
energy spectrum of the $B\to X_s\gamma$ decay \cite{ag}, which provides the
information of the $b$ quark mass $M_b$ and of the HQET parameter
$\lambda_1$ for the $B_d$ meson. Based on this information, we propose
reasonable $B_s$ meson and $\Lambda_b$ baryon distribution functions by
assuming that they correspond to the same $\lambda_1$. Convoluting these
distribution functions with the perturbative factors, we predict the
correct lifetimes of the $b$-hadrons: $\tau(B_d)=1.56$ ps, $\tau(B_s)=1.46$
ps and $\tau(\Lambda_b)=1.22$ ps, and thus the correct ratios
$\tau(B_s)/\tau(B_d)=0.94$ and $\tau(\Lambda_b)/\tau(B_d)=0.78$. By varying
the $B$ meson distribution function slightly, we obtain the $B_u$ meson
lifetime $\tau(B_u)=1.62$ ps, and the ratio $\tau(B_u)/\tau(B_d)=1.04$. It
will be shown that perturbative contributions play an essential role in the
explanation of the $b$-hadron lifetimes, and that our predictions are
insensitive to the variation of the distribution functions, as expected
from the quark-hadron duality. Our results of the semileptonic branching
ratio $B_{\rm SL}=B(B\to Xl{\bar\nu})=10.16\%$ and of the average charm
yield $\langle n_c\rangle=1.17$ per $B$ decay are also consistent with
experimental data. We present the semileptonic branching ratios and the
charm yields in $B_s$ meson and $\Lambda_b$ baryon decays as well, which
can be tested in future experiments.

In Sec.~II the factorization theorem for semileptonic $b$-hadron decays is
constructed. The equivalence of the $B$ meson distribution function to that
derived in the HQET framework is demonstrated in Sec.~III. The three-scale
factorization theorem for nonleptonic decays is developed in Sec.~IV.
Various logarithmic corrections are summed to all orders using the
resummation technique and RG equations in Sec.~V. In Sec.~VI we present
the factorization formulas for the $b$-hadron decay widths and their
numerical analysis. Section VII is the conclusion. The
appendix contains the detailed derivation of the allowed phase space for
heavy hadron decays.

\section{FACTORIZATION OF SEMILEPTONIC DECAYS}

We start with the factorization of the simplest case, the semileptonic
decays $B\to X_cl{\bar \nu}$, which correspond to the $b\to cl{\bar\nu}$
decays at the quark level. These decays involve only the $B$ meson
distribution function and the hard $b$ quark decay amplitudes due to the 
lack of the characteristic scale of the $W$ boson mass. Nonperturbative 
dynamics is
reflected by infrared poles in radiative corrections to quark-level
amplitudes in perturbation theory. According to PQCD factorization theorems,
these poles are absorbed into a hadron distribution function, which must be
derived by nonperturbative methods or extracted from experimental data.
A distribution function, being universal, is determined once for all,
and then employed to make predictions for other processes containing the
same hadron. In this section we demonstrate how to isolate infrared poles
from radiative corrections and factorize them into the $B$ meson
distribution function.

Consider the one-loop corrections to the $b\to cl{\bar\nu}$ decays
shown in Fig.~1. Because both the $b$ and $c$ quarks are massive,
there are no collinear (mass) divergences, and we concentrate only on
soft divergences from vanishing loop momenta. The self-energy correction to
the $b$ quark in Fig.~1(a) is written as
\begin{eqnarray}
\Sigma^{(a)} u_b=-ig^2C_F\mu^{2\epsilon}
\int\frac{d^{4-2\epsilon}l}{(2\pi)^{4-2\epsilon}}
\frac{1}{l^2}\frac{{\not p}_b+M_b}{p_b^2-M_b^2}\gamma_\mu
\frac{{\not p}_b-\not l+M_b}{(p_b-l)^2-M_b^2}\gamma^\mu u_b
\delta(p_c^2-M_c^2)\;,
\label{a1}
\end{eqnarray}
with $C_F=4/3$ a color factor, $p_b$ and $u_b$ the $b$ quark momentum and 
spinor, respectively, and $p_c=p_b-q$ and $M_c$ the $c$ quark momentum 
and mass, respectively, $q$ being the lepton pair momentum. The 
$\delta$-function from the final-state cut on the outgoing $c$ quark is 
included for the discussion of the soft cancellation between
virtual and real corrections below. The $b$ quark propagator
$({\not p}_b+M_b)/(p_b^2-M_b^2)$ after the self-energy correction helps the
extraction of the soft pole in Eq.~(\ref{a1}).

The soft divergence of $\Sigma^{(a)}$ is isolated by the eikonal ($l\to 0$)
approximation,
\begin{eqnarray}
\Sigma^{(a)}_{\rm soft} u_b=-ig^2C_F\mu^{2\epsilon}
\int\frac{d^{4-2\epsilon}l}{(2\pi)^{4-2\epsilon}}
\frac{1}{l^2}\frac{{\not p}_b+M_b}{p_b^2-M_b^2}
\gamma_\mu\frac{2p_b^\mu}{-2p_b\cdot l+p_b^2-M_b^2} u_b
\delta(p_c^2-M_c^2)\;,
\label{1a1}
\end{eqnarray}
where the term $\not l$ in the numerator and $l^2$ in the denominator of
the $b$ quark propagator have been neglected, and the equation of motion
$({\not p}_b-M_b)u_b=0$ has been applied to obtain the factor $2p_b^\mu$.
We make the following expansion because of the on-shell condition
$p_b^2-M_b^2\to 0$:
\begin{equation}
\frac{1}{-2p_b\cdot l+p_b^2-M_b^2}
=-\frac{1}{2p_b\cdot l}-\frac{p_b^2-M_b^2}{4(p_b\cdot l)^2}\;,
\end{equation}
where the first term on the right-hand side does not contribute, since it
leads to an integrand with an odd power in $l$. The numerator of the second
term removes the on-shell pole of the $b$ quark propagator after the
self-energy correction. Employing the equation of motion again, we obtain
\begin{eqnarray}
\Sigma^{(a)}_{\rm soft}=ig^2C_F\mu^{2\epsilon}\int\frac{d^{4-2\epsilon}l}
{(2\pi)^{4-2\epsilon}}\frac{v^2}{l^2(v\cdot l)^2}\delta(p_c^2-M_c^2)\;,
\label{eik}
\end{eqnarray}
where $v\equiv p_b/M_b$ is the $b$ quark velocity.
Obviously, Eq.~(\ref{eik}) does not depend on  the spin and mass of the $b$
quark.

The loop integral for Fig.~1(b) with a real gluon attaching the $b$ quarks
before and after the final-state cut is written as
\begin{eqnarray}
u_b\Sigma^{(b)}u_b&=&-g^2C_F\mu^{2\epsilon}
\int\frac{d^{4-2\epsilon}l}{(2\pi)^{4-2\epsilon}}
u_b\gamma_\mu\frac{{\not p}_b-\not l+M_b}{(p_b-l)^2-M_b^2}
\Gamma\frac{{\not p}_b-\not l+M_b}{(p_b-l)^2-M_b^2}\gamma^\mu 
\nonumber\\
& &\times 2\pi\delta(l^2)\delta((p_c-l)^2-M_c^2)\;,
\label{b1}
\end{eqnarray}
where $\Gamma$ represents the other irrelevant vertices and propagators.
The $\delta$-function is associated with the final-state $c$ quark with the
momentum $p_c-l=p_b-q-l$ in the case of real gluon emission. The soft
divergence in Eq.~(\ref{b1}) is also extracted by applying the eikonal
approximation and the equation of motion to the two $b$ quark propagators,
leading to
\begin{eqnarray}
\Sigma^{(b)}_{\rm soft}=-g^2C_F\mu^{2\epsilon}
\int\frac{d^{4-2\epsilon}l}{(2\pi)^{4-2\epsilon}}
\frac{v^2}{(v\cdot l)^2}2\pi\delta(l^2)\delta((p_c-l)^2-M_c^2)\;.
\label{rel}
\end{eqnarray}
The above expression indicates that under the eikonal approximation, the 
$b$ quark
propagator is replaced by the eikonal propagator $1/(v\cdot l)$ and the
quark-gluon vertex $\gamma^\mu$ is replaced the eikonal vertex $v^\mu$.
These Feynman rules for an eikonal line are exactly the same as those
associated with the large component $h_v(x)$ of the rescaled $b$ quark
field $b_v(x)$,
\begin{equation}
h_v(x)=\frac{1+\not v}{2}b_v(x)\;,\;\;\;\;
b_v(x)=\exp(iM_bv\cdot x)b(x)\;,
\label{hq}
\end{equation}
that appears in HQET. This observation will become essential, as we
demonstrate the identity of the $B$ meson distribution functions constructed
in the PQCD factorization theorems and in HQET.

A straightforward calculation gives
\begin{eqnarray}
\Sigma^{(a)}_{\rm soft}&=&\frac{\alpha_s}{2\pi}C_F\frac{1}{-\epsilon}
\delta(p_c^2-M_c^2)\;,
\label{1as}\\
\Sigma^{(b)}_{\rm soft}&=&-\frac{\alpha_s}{\pi}C_F
\frac{(\mu^2 p_c^2)^\epsilon}{(p_c^2-M_c^2)^{1+2\epsilon}}\;.
\label{1bs}
\end{eqnarray}
The ultraviolet pole in $\Sigma^{(a)}_{\rm soft}$ has been subtracted, 
$1/(-\epsilon)$ with $\epsilon < 0$ denotes the soft pole, and the other
finite terms irrelevant to our discussion have been dropped. The real
gluon correction $\Sigma^{(b)}_{\rm soft}$ in fact contains a soft pole as
$p_c^2\to M_c^2$. Following the standard factorization of deep inelastic
scattering, this pole is extracted through the convolution with a $B$
meson structure function. The introduction of such a structure
function implies that the $b$ quark momentum $p_b$ is allowed to vary.
A variable $p_b$ is natural, since the light degrees of freedom in the
$B$ meson share various amount of meson momentum, such that the $b$ quark 
is not always at rest.

Define the residual momentum of the $b$ quark as $k=p_b-M_bv$, 
$v=(v^+,v^-,{\bf v}_\perp)=(1,1,{\bf 0})/\sqrt{2}$ in the light-cone
notation, for which $k=(k^+,0,{\bf 0})$ is a convenient parametrization.
The $B$ meson structure function $f(k^+)$ then describes the probability of
finding a $b$ quark with the residual momentum $k^+$ inside the $B$ meson.
Consequently, the one-loop corrections in Figs.~1(a) and 1(b) are modified
into the convolutions of Eqs.~(\ref{a1}) and (\ref{b1}) with $f(k^+)$,
respectively, where the $c$ quark momentum is replaced by $p_c=M_bv+k-q$.
It is legitimate to reexpress the factor $1/(p_c^2-M_c^2)^{1+2\epsilon}$ in
Eq.~(\ref{1bs}) as
\begin{eqnarray}
\frac{1}{(p_c^2-M_c^2)^{1+2\epsilon}}
=\frac{1}{-2\epsilon}\delta(p_c^2-M_c^2)
+\frac{1}{(p_c^2-M_c^2)_+}\;.
\label{ind0}
\end{eqnarray}
The second term in the above expression is defined via the convolution
with an arbitrary function $F(k^+)$,
\begin{eqnarray}
\int dk^+\frac{F(k^+)}{(p_c^2-M_c^2)_+}
=\int dk^+\frac{F(k^+)-F({\bar k})}{p_c^2-M_c^2}\;,
\label{ind}
\end{eqnarray}
in which
\begin{eqnarray}
{\bar k}=-\frac{M_b^2-2M_bq^0+q^2-M_c^2}{2(M_bv^--q^-)}\;,
\end{eqnarray}
is the value of $k^+$ determined by the on-shell condition $p_c^2=M_c^2$.
Apparently, Eq.~(\ref{ind}) is infrared finite as $k^+\to {\bar k}$
($p_c^2\to M_c^2$), and thus the second term in Eq.~(\ref{ind0}) is
negligible. The first term in Eq.~(\ref{ind0}) leads to
\begin{eqnarray}
\Sigma^{(b)}_{\rm soft}=-\frac{\alpha_s}{2\pi}C_F
\frac{1}{-\epsilon}\delta(p_c^2-M_c^2)\;,
\label{sib}
\end{eqnarray}
which cancels $\Sigma^{(a)}_{\rm soft}$ in Eq.~(\ref{1as}).

A similar cancellation occurs between the self-energy correction to the $c$
quark in Fig.~1(c) and its corresponding real gluon correction in Fig.~1(d),
when they are convoluted with the $B$ meson structure function. The loop
integrals are the same as those in Eqs.~(\ref{eik}) and (\ref{rel}) but with
the $b$ quark velocity $v$ replaced by the $c$ quark velocity 
$v_c\equiv p_c/M_c$.
Using the eikonal approximation, the soft structure of the vertex correction
in Fig.~1(e) with the virtual gluon attaching the $b$ and $c$ quarks is
given by
\begin{eqnarray}
\Sigma^{(e)}_{\rm soft}=-ig^2C_F\mu^{2\epsilon}
\int\frac{d^{4-2\epsilon}l}{(2\pi)^{4-2\epsilon}}
\frac{v\cdot v_c}{l^2 v\cdot l v_c\cdot l}\delta(p_c^2-M_c^2)\;.
\end{eqnarray}
The soft structure extracted from the corresponding real correction
in Fig.~1(f) is
\begin{eqnarray}
\Sigma^{(f)}_{\rm soft}=g^2C_F\mu^{2\epsilon}
\int\frac{d^{4-2\epsilon}l}{(2\pi)^{4-2\epsilon}}
\frac{v\cdot v_c}{v\cdot l v_c\cdot l}2\pi\delta(l^2)
\delta((p_c-l)^2-M_c^2)\;.
\end{eqnarray}
Following the similar reasoning, we show that $\Sigma^{(e)}_{\rm soft}$ and
$\Sigma^{(f)}_{\rm soft}$ cancel exactly. The above soft cancellation can be
generalized to all orders straightforwardly using the eikonal approximation,
under which soft gluons detach from the quarks. In conclusion, the exact
soft cancellation between virtual and real corrections to $b$ quark
decays must be implemented by introducing the $B$ meson structure
function. This is consistent with the Kinoshita-Lee-Nauenberg theorem,
which states that infrared divergences in radiative corrections to a QCD
process cancel when both final and initial states are summed. The $B$ meson
structure function $f(k^+)$ simply provides a weighting factor for the
summation over the initial $b$ quark states.

The one-loop contributions to the hard $b$ quark decay amplitude $H$ for the
semileptonic decays are defiined as the difference of the full radiative
corrections and their eikonalized versions, that have been absorbed into the
$B$ meson structure function. As a demonstration, we perform the
factorization of the self-energy correction to the $b$ quark:
\begin{eqnarray}
& &({\rm tree}\;\;{\rm diagram})+{\rm Fig.~1(a)}
\nonumber\\
&=&({\rm tree}\;\;{\rm diagram})+{\rm Fig.~1(a)}
-({\rm tree}\;\;{\rm diagram})\times \Sigma^{(a)}_{\rm soft}
+({\rm tree}\;\;{\rm diagram})\times \Sigma^{(a)}_{\rm soft}\;,
\nonumber\\
&=&[({\rm tree}\;\;{\rm diagram})+{\rm Fig.~1(a)}
-({\rm tree}\;\;{\rm diagram})\times \Sigma^{(a)}_{\rm soft}]\times
(1+\Sigma^{(a)}_{\rm soft})+O(\alpha_s^2)\;.
\label{fas}
\end{eqnarray}
The first and second factors in the last line of the above expression
belong to the first-order $H$ and $f(k^+)$, respectively. Extending the 
above procedures to all orders, we derive the factorization formula for the
semileptonic decays, which is a convolution of the hard amplitude with the
$B$ meson structure function. Because of the soft subtraction
$({\rm tree}\;\;{\rm diagram})\times \Sigma^{(a)}_{\rm soft}$, $H$ is
infrared finite and calculable at the quark level in perturbation theory.

\section{MOMENTS OF THE DISTRIBUTION FUNCTION}

A formal definition of the $B$ meson structure function $f(k^+)$ should 
reflect the soft dynamics associated with the $B$ meson. Hence, we replace
the $b$ quark line by an eikonal line in the direction $v$, which collects
infinite many soft gluons radiated by the $b$ quark. From Eq.~(\ref{hq}),
the dependence on the $b$ quark mass $M_b$ is removed by employing the
rescaled $b$ quark field $b_v$. Specifying the large component $h_v$ of
$b_v$, the spin of the $b$ quark remains fixed as radiating gluons, and its
dependence also decouples. Therefore, the propagator and the gluon vertex
associated with $h_v$ in HQET are exactly the same as the eikonal propagator
and vertex, implying that $f(k^+)$ is defined in terms of $h_v$. The initial
and final $b$ quarks have a separation $y^-$ in the minus coordinate, since
the $b$ quark momentum varies in the plus direction. At last, to render the
definition gauge invariant, we insert a path-ordered exponential
$P\exp[i\int_0^{y^-}dtn\cdot A(tn)]$ in between the initial and final $b$
quark fields, with $n=(0,1,{\bf 0})$ a vector on the light cone, forming
\begin{equation}
f(k^+)=\int\frac{dy^-}{2\pi}e^{ik^+y^-}
\langle B(v)|{\bar h}_v(0)P\exp\left[-i\int_0^{y^-}dtn\cdot A(tn)\right]
h_v(y^-)|B(v)\rangle\;,
\label{deb}
\end{equation}
where $|B(v)\rangle$ is the initial state of the $B$ meson. 
The Feynman rules for the
path-ordered exponential are an eikonal line in the direction $n$ with the
propagator $1/(n\cdot l)$ and the vertex $n^\mu$.

The effects of the variation of the $b$ quark momentum from the mass
shell, determined by $f(k^+)$, are
nonperturbative. Because the variation is of order
$\Lambda_{\rm QCD}$, it is appropriate to expand the nonperturbative
corrections contained in $f(k^+)$ in terms of its moments \cite{N}:
\begin{eqnarray}
& &\int dk^+ f(k^+)=\langle B(v)|{\bar h}_v(0)h_v(0)|B(v)\rangle=1\;,
\label{mo0}\\
& &\int dk^+ k^+ f(k^+)=
\langle B(v)|{\bar h}_v(0)iD^+ h_v(0)|B(v)\rangle=0\;,
\label{mo1}\\
& &\int dk^+ k^{+2} f(k^+)=
\langle B_v|{\bar h}_v(0)(iD^+)^2h_v(0)|B(v)\rangle\equiv
-\lambda_1/6\;,
\label{mo2}
\end{eqnarray}
where Eq.~(\ref{mo0}) is the normalization of $f(k^+)$, Eq.~(\ref{mo1})
is the consequence of the equation of motion for $h_v$, and Eq.~(\ref{mo2})
defines the HQET parameter $\lambda_1$. Equation (\ref{mo1}) implies that
the $O(1/M_b)$ nonperturbative correction to the semileptonic decays does
not exist, as observed in HQET. That is, the quark-hadron duality, which
states the equality of heavy hadron decay widths to heavy quark decay
widths up to small difference of $O(1/M_b^2)$ from $\lambda_1$, is respected
in the PQCD formalism.

As argued before, the $B$ meson structure function allows a variable $b$
quark momentum, whose source is the effect of the light degrees of freedom
in the $B$ meson. This effect demands the consideration of $B$ meson,
instead of $b$ quark, decays, and thus the $B$ meson kinematics. Therefore,
we transform the structure function $f(k^+)$ into the usual $B$ meson
distribution function $f_B(z)$, $z=p_b^+/P_B^+$, which describes the
probability to find a $b$ quark with the plus momentum $zP_B^+$ \cite{N}.
The relation between $f(k^+)$ and $f_B(z)$ cab be extracted from their
moments by employing the variable transformation,
\begin{equation}
k^+=p_b^+-M_bv^+=\frac{zM_B-M_b}{\sqrt{2}}\;.
\label{var}
\end{equation}
Equation (\ref{mo0}) leads to the normalization of $f_B(z)$,
\begin{equation}
\int_0^1 dz f_B(z)=1\;,\;\;\;\;
f_B(z)\equiv \frac{M_B}{\sqrt{2}}f\left(\frac{zM_B-M_b}
{\sqrt{2}}\right)\;.
\label{nor}
\end{equation}
Equation (\ref{mo1}) gives
\begin{eqnarray}
\int dz (1-z) f_B(z)=\left(1-\frac{M_b}{M_B}\right)\int dz f_B(z)
=\frac{{\bar\Lambda}}{M_B}
+O\left(\frac{\Lambda^2_{\rm QCD}}{M_B^2}\right)\;,
\label{ke}
\end{eqnarray}
into which Eq.~(\ref{nor}) has been inserted. Applying a similar manipulation
to Eq.~(\ref{mo2}), we arrive at
\begin{equation}
\int dz (1-z)^2 f_B(z)=\frac{1}{M_B^2}\left({\bar\Lambda}^2-
\frac{\lambda_1}{3}\right)
+O\left(\frac{\Lambda^3_{\rm QCD}}{M_B^3}\right)\;.
\label{ke1}
\end{equation}
The HQET parameters ${\bar\Lambda}$ and $\lambda_1$ satisfy the relation
\begin{eqnarray}
{\bar\Lambda}=M_B-M_b+\frac{\lambda_1}{2M_b}\;.
\label{mass}
\end{eqnarray}
Therefore, the heavy hadron kinematics is introduced into the PQCD
formalism rigorously via Eqs.~(\ref{deb})-(\ref{ke1}).
It is then clear that the PQCD approach is equivalent to HQET in the
treatment of nonperturbative corrections, and the moment
${\bar\Lambda}/M_B$ is attributed to the replacement of the $b$ quark mass
by the $B$ meson mass. An advantage of the $B$ meson kinematics is that it
provides the correct kinematic bounds. Take the lepton energy spectrum as
an example. Using the $b$ quark kinematics, the maximal lepton energy
$M_b/2$ is smaller than the correct value $M_B/2$. 

The $B$ meson distribution function must be determined by means outside the
PQCD regime. For its functional form, we propose \cite{ab}
\begin{equation}
f_B(z)=N\frac{z(1-z)^2}{[(z-a)^2+\epsilon z]^2}\;,
\label{eq: aub}
\end{equation}
where the three parameters $N$, $a$ and $\epsilon$ are obtained from
the best fit to experimental data. We have extracted $f_B(z)$ from the
photon energy spectrum of the radiative decay $B\to X_s\gamma$, given by
\cite{ag}
\begin{equation}
f_B(z)= \frac{0.02647 z(1-z)^2}{[( z - 0.95 )^2 + 0.0034 z]^2}\;.
\label{bdf}
\end{equation}
The maximum of $f_B$ occurs at $z\sim 1$ as expected, since the $b$ quark 
carries most
of the $B$ meson momentum. Substituting the above expression into
Eqs.~(\ref{nor}), (\ref{ke}) and (\ref{ke1}), which relate the three
parameters $N$, $a$ and $\epsilon$ to the HQET parameters, we obtain
${\bar\Lambda}=0.65$ GeV and $\lambda_1=-0.71$ GeV$^2$.

As to the distribution functions for other $b$-hadrons, we assume the
parametrizations of the $B_s$ meson and $\Lambda_b$ baryon distribution
functions, $f_{B_s}(z)$ and $f_{\Lambda_b}(z)$, respectively, the same as
Eq.~(\ref{eq: aub}). The moments of $f_{B_s(\Lambda_b)}$ need to be
treated as free parameters, since they have not been determined yet. The
relations of the parameters $N$, $a$ and $\epsilon$ to the first three
moments of $f_{B_s(\Lambda_b)}$ are similar to Eqs.~(\ref{nor}), (\ref{ke})
and (\ref{ke1}):
\begin{eqnarray}
& &\int_0^1 f_{B_s(\Lambda_b)}(z) dz=1\;,
\nonumber\\
& &\int_0^1 (1-z)f_{B_s(\Lambda_b)}(z) dz=
\frac{{\bar\Lambda}_{B_s(\Lambda_b)}}
{M_{B_s(\Lambda_b)}}+O(\Lambda^2_{\rm QCD}/M_{B_s(\Lambda_b)}^2)\;,
\nonumber \\
& &\int_0^1 (1-z)^2f_{B_s(\Lambda_b)}(z) dz=\frac{1}{M^2_{B_s(\Lambda_b)}}
\left({\bar\Lambda}^2_{B_s(\Lambda_b)}
-\frac{1}{3}\lambda_{1B_s(\Lambda_b)}\right)
+O(\Lambda^3_{\rm QCD}/{M_{B_s(\Lambda_b)}^3}),
\label{eq: ap}
\end{eqnarray}
with
\begin{eqnarray}
& &\lambda_{1B_s(\Lambda_b)}=-6
\langle B_s(\Lambda_b)|{\bar b}(iD_\perp)^2b|B_s(\Lambda_b)\rangle
\nonumber\\
& &M_{B_s(\Lambda_b)}=M_b+{\bar\Lambda}_{B_s(\Lambda_b)}-
\frac{\lambda_{1B_s(\Lambda_b)}}{2M_b}\;.
\label{eq: ar}
\end{eqnarray}

Note that only $\lambda_{1B_s}$ and $\lambda_{1\Lambda_b}$ are free 
parameters. Because of the parameters ${\bar\Lambda}=0.65$ GeV and
$\lambda_1=-0.71$ GeV$^2$ and $M_B=5.279$ GeV, the $b$ quark mass
$M_b=4.551$ GeV is fixed by Eq.~(\ref{mass}). Substituting this $M_b$ into
Eq.~(\ref{eq: ar}) with $M_{B_s}=5.369$ GeV and $M_{\Lambda_b}=5.621$ GeV,
${\bar\Lambda}_{B_s}$ and ${\bar\Lambda}_{\Lambda_b}$ will be derived,
if $\lambda_{1B_s}$ and $\lambda_{1\Lambda_b}$ are chosen. Using the values
of ${\bar\Lambda}_{B_s(\Lambda_b)}$ and $\lambda_{1B_s(\Lambda_b)}$, 
combined with the normalization of $f_{B_s(\Lambda_b)}$, we obtain the
corresponding parameters $N$, $a$ and $\epsilon$. In the numerical analysis
we shall assume that the HQET parameters $\lambda_1$ are the same
for all $b$-hadrons, {\it i.e.},
$\lambda_{1B_s}=\lambda_{1\Lambda_b}=\lambda_1=-0.71$ GeV$^2$, which
correspond to ${\bar\Lambda}_{B_s}=0.74$ GeV and
${\bar\Lambda}_{\Lambda_b}=0.99$ GeV. The $B_s$ meson and $\Lambda_b$ baryon
distribution functions are then given by
\begin{eqnarray}
f_{B_s}(z)&=& \frac{0.02279 z(1-z)^2}{[( z - 0.93 )^2 + 0.0043 z]^2}\;,
\label{bds}\\
f_{\Lambda_b}(z)&=& \frac{0.02095 z(1-z)^2}{[( z - 0.89 )^2 + 0.0068 z]^2}\;.
\label{bdb}
\end{eqnarray}

For the $B_u$ meson distribution function, the reasoning is a bit different.
In fact, we can not distinguish the $B_u$ and $B_d$ meson distribution
functions, if $f_B$ is extracted from the photon energy spectrum of the
$B\to X_s \gamma$ decay, which acquires contributions from both
mesons \cite{C}. To explain the $B_u$ meson lifetime in Sec.~VI, we
shall vary $f_B$ slightly:
\begin{eqnarray}
f_{B_u}(z)= \frac{0.03358 z(1-z)^2}{[( z - 0.96 )^2 + 0.0034 z]^2}\;,
\label{bdu}
\end{eqnarray} 
which corresponds to ${\bar\Lambda}_{B_u}=0.65$ GeV and
$\lambda_{1B_u}=-0.81$ GeV$^2$, about 10\% deviation from $\lambda_1$ for
the $B_d$ meson.

\section{FACTORIZATION OF NONLEPTONIC DECAYS}

The factorization theorem for nonleptonic $B$ meson decays is more
complicated than the semileptonic case. Nonleptonic decays involve three
scales: the $W$ boson mass $M_W$, at which the matching condition of the
effective Hamiltonian to the full Hamiltonian is defined, the characteristic
scale $t$, which reflects the specific dynamics of a decay mode, and the $b$
quark transverse momentum $p_\perp$, which serves as a factorization scale.
Below the factorization scale, dynamics is regarded as being purely
nonperturbative, and absorbed into the $B$ meson structure function, if it
is soft, or into jet functions associated with light energetic final-state
quarks, if it is collinear. Above the factorization scale, PQCD is reliable,
and radiative corrections are absorbed into a hard $b$ quark decay amplitude
characterized by $t$, or into a harder function characterized by $M_W$. In
this section we shall demonstrate how to construct the three-scale
factorization theorem for nonleptonic $B$ meson decays \cite{ah}.

Consider a nonleptonic $b$ quark decay $b\to cq{\bar q}'$ through a $W$
boson emission up to $O(\alpha_{s})$ virtual gluon corrections. The sum of
the full diagrams does not possess ultraviolet divergences because of the
current conservation and the presence of the $W$ boson propagator. Reexpress
a full diagram into two terms, where the first term is obtained by shrinking
the $W$ boson line into a point, and the second term is the difference of
the full diagram and the first term. The first term corresponds to a local
four-fermion operator $({\bar q} q')({\bar c} b)$ appearing in the full or
effective Hamiltonian with $({\bar q} q')={\bar q} \gamma_\mu(1-\gamma_5)q'$
the $V-A$ current, and is absorbed into a decay amplitude $H'(p_b,p_j,\mu)$,
$p_j$ being the outgoing quark momenta. $H'$ is characterized by momenta
smaller than the $W$ boson mass $M_W$, since gluons in $H'$ do not ``see"
the $W$ boson. Note that the factorization in $H'$, which still contains the
contributions characterized by the hadronic scale, is not complete yet. The
second term, characterized by momenta of order $M_W$ due to the subtraction
term, is absorbed into the harder function $H_r(M_W,\mu)$. Following the
steps that lead to Eq.~(\ref{fas}), we derive the
factorization formula for the $b$ quark decay width \cite{ah},
\begin{equation}
\Gamma=H_r(M_W,\mu)\times H'(p_b,p_j,\mu)\;,
\label{fa1}
\end{equation}
Though the full diagrams are ultraviolet finite, the factorization
introduces the dependence on the renormalization scale $\mu$ into $H_r$ and
$H'$, because the diagrams with the $W$ boson line shrunk are divergent.

We then investigate infrared divergences from the diagrams contained in
$H'$. These diagrams are basically similar to those in Fig.~1, where the
four-fermion vertices have been adopted, and the final-state quark may
represent the $c$, $q$, or ${\bar q}'$ quark. At this stage real gluon
corrections are taken into account. If the final-state quark is $c$, the
discussion the same as in Sec.~II leads to the $B$ meson structure function
defined by Eq.~(\ref{deb}), which is thus universal. If the final-state
quark is $q$, the loop integrals for Figs.~1(c) and 1(d) are written as
\begin{eqnarray}
({\not p}_j+m)\Sigma^{'(c)}&=&-ig^2C_F\mu^{2\epsilon}
\int\frac{d^{4-2\epsilon}l}{(2\pi)^{4-2\epsilon}}({\not p}_j+m)
\gamma_\mu\frac{{\not p}_j-\not l+m}{(p_j-l)^2-m^2}\gamma^\mu
\frac{{\not p}_j+m}{p_j^2-m^2}
\frac{1}{l^2}
\nonumber\\
& &\times \delta(p_j^2-m^2)\;,
\label{3a}\\
({\not p}_j+m)\Sigma^{'(d)}&=&-g^2C_F\mu^{2\epsilon}
\int\frac{d^{4-2\epsilon}l}{(2\pi)^{4-2\epsilon}}
\frac{{\not p}_j+m}{p_j^2-m^2}
\gamma_\mu({\not p}_j-\not l+m)\gamma^\mu\frac{{\not p}_j+m}{p_j^2-m^2}
\nonumber\\
& &\times 2\pi\delta(l^2)\delta((p_j-l)^2-m^2)\;,
\label{3b}
\end{eqnarray}
with $p_j$ and $m$ the final-state quark momentum and mass, respectively.

Performing the loop integrations directly, we obtain
\begin{eqnarray}
\Sigma^{'(c)}&=&\frac{\alpha_s}{2\pi}C_F\left(\frac{1}{-\epsilon}
-\frac{3}{2}\ln\frac{\mu^2}{m^2}\right)\delta(p_j^2-m^2)\;,
\label{3a3}\\
\Sigma^{'(d)}&=&-\frac{\alpha_s}{\pi}C_F
\frac{(\mu^2p_j^2)^\epsilon}{(p_j^2-m^2)^{1+2\epsilon}}\;.
\label{3b3}
\end{eqnarray}
The terms finite as $m\to 0$ have been dropped and the ultraviolet pole has
been subtracted in Eq.~(\ref{3a3}). Similarly, Eq.~(\ref{3b3}) is
reexpressed, following Eq.~(\ref{ind0}), as
\begin{eqnarray}
\Sigma^{'(d)}=-\frac{\alpha_s}{2\pi}C_F\frac{1}{-\epsilon}
\delta(p_j^2-m^2)\;.
\label{3b2}
\end{eqnarray}
Obviously, the soft poles $1/(-\epsilon)$ disappear in the sum of
Eqs.~(\ref{3a3}) and (\ref{3b2}) as expected. For $m\not= 0$, there are no
other infrared divergences, the same as in the $c$ quark case. However, an
additional infrared divergence, {\it i.e.}, the collinear divergence,
appears when the invariant mass $p_j^2$ of the final-state quark is equal
to $m^2$ and vanishes, which remains even in the sum of Eqs.~(\ref{3a3}) and
(\ref{3b2}). This divergence comes from the region with the loop momentum
$l$ being parallel to $p_j$ and not small, for which both $(p_j-l)^2$ and
$l^2$ approach zero. Note that $p_j^2$ vanishes only in the kinematic
end-point region. To absorb this end-point singularity, we introduce a jet
function associated with the energetic outgoing light quark. Figures 1(c)
and 1(d) are then trivially factorized into the jet function.

For Figs.~1(e) and 1(f) with a gluon attaching the $b$ quark and the
$q$ quark, the discussion is the same as for the $c$ quark
case, if the $q$ quark is massive: there are no collinear divergences, and
soft divergences cancel, as the two diagrams are convoluted with the $B$
meson structure function. If the $q$ quark is massless,
the loop integrals for Figs.~1(e) and 1(f) are
written as
\begin{eqnarray}
{\not p}_j\Sigma^{'(e)}&=&-ig^2C_F\mu^{2\epsilon}
\int\frac{d^{4-2\epsilon}l}{(2\pi)^{4-2\epsilon}}{\not p}_j
\gamma_\mu\frac{{\not p}_j-\not l}{(p_j-l)^2}
\Gamma_4\frac{{\not p}_b-\not l+M_b}{(p_b-l)^2-M_b^2}\gamma^\mu
\frac{1}{l^2}\delta(p_j^2)\;,
\label{3c}\\
{\not p}_j\Sigma^{'(f)}&=&-g^2C_F\mu^{2\epsilon}
\int\frac{d^{4-2\epsilon}l}{(2\pi)^{4-2\epsilon}}
\frac{{\not p}_j}{p_j^2}\gamma_\mu({\not p}_j-\not l)
\Gamma_4\frac{{\not p}_b-\not l+M_b}{(p_b-l)^2-M_b^2}\gamma^\mu
2\pi\delta(l^2)\delta((p_j-l)^2)\;,
\label{3d}
\end{eqnarray}
where $\Gamma_4$ represents the four-fermion vertex. Similarly, collinear
divergences from $l$ parallel to $p_j$ with $p_j^2\to 0$ exist and remain 
after summing
$\Sigma^{'(e)}$ and $\Sigma^{'(f)}$. These collinear divergences should be
absorbed into the jet function associated with the $q$ quark.

The factorization of Figs.~1(e) and 1(f) in the collinear region requires
an eikonal approximation. For convenience, we assume that $p_j$ is in the
plus direction at the kinematic end point. When $l$ is parallel to $p_j$,
only $\gamma_-$ in the gamma matrices $\gamma_\mu$ contributes, indicating
that only $\gamma^-$ in $\gamma^\mu$ contributes, and that $\not l$ in the
numerator of the $b$ quark propagator is negligible. Using the equation of
motion for the $b$ quark spinor, we have the eikonal approximation,
\begin{equation}
\frac{{\not p}_b+M_b}{(p_b-l)^2-M_b^2}\gamma^\mu
=\frac{2p_b^\mu}{l^2-2p_b\cdot l}
\approx\frac{p_b^-\delta^{-\mu}}{-p_b\cdot l}
=-\frac{n^\mu}{n\cdot l}\;.
\label{jfa}
\end{equation}
That is, the collinear gluon decouples from the $b$ quark, and attaches
the eikonal line in the direction $n$ defined in Sec.~II. Using
Eq.~(\ref{jfa}), Eqs.~(\ref{3c}) and (\ref{3d}) reduce to
\begin{eqnarray}
\Sigma^{'(e)}_{\rm coll}&=&ig^2C_F\mu^{2\epsilon}
\int\frac{d^{4-2\epsilon}l}{(2\pi)^{4-2\epsilon}}
\gamma_\mu\frac{{\not p}_j-\not l}{(p_j-l)^2}
\frac{n^\mu}{n\cdot l}\frac{1}{l^2}\delta(p_j^2)\;,
\label{3c1}\\
\Sigma^{'(f)}_{\rm coll}&=&g^2C_F\mu^{2\epsilon}
\int\frac{d^{4-2\epsilon}l}{(2\pi)^{4-2\epsilon}}
\frac{1}{p_j^2}\gamma_\mu({\not p}_j-\not l)
\frac{n^\mu}{n\cdot l}2\pi\delta(l^2)\delta((p_j-l)^2)\;.
\label{3d1}
\end{eqnarray}

Since the eikonal approximation in Eq.~(\ref{jfa}) also holds for a
massless quark propagator with $M_b=0$, the factorization of the other set
of diagrams with a gluon attaching the $q$ and ${\bar q}'$ quarks in the
collinear region is similar: the collinear gluon decouples from the
${\bar q}'$ quark and attaches the eikonal line in the direction $n$.
Extending the above analysis to all orders, the jet function is defined
as the collection of infinite many gluon exchanges among the $q$ quark and
the eikonal lines in the direction $n$ on both sides of the final-state cut.
Note that the self-energy corrections to the eikonal line are excluded.
The above conclusion applies to the other final-state ${\bar q}'$ quark:
if the ${\bar q}'$ quark is massless, there exist collinear divergences
at the kinematic end point, and another jet function needs to be introduced.

With the jet functions and the $B$ meson structure function
defined in Eq.~(\ref{deb}), all the infrared divergences in nonleptonic
$B$ meson decays are extracted and factorized appropriately. The hard $b$
quark decay amplitude $H(t,\mu)$ is then defined as the difference of
$H'$ and those diagrams which have been absorbed into the jet functions and
the $B$ meson structure funciton, and thus calculable in perturbation
theory. Following Eq.~(\ref{fas}), we obtain the factorization
of $H'$,
\begin{equation}
H'(p_b,p_j,\mu)=H(t,\mu)\times \prod_jJ_j(p_j,\mu) \times f(k^+,\mu)\;,
\label{fa2}
\end{equation}
where the index $j$ runs over the light final-state quarks. Note that a
trace of $H$ and $J_j$ is necessary, since the jet function carries the spin
structure of the corresponding final-state quark. Combining Eqs.~(\ref{fa1})
and (\ref{fa2}), the three-scale factorization formula for nonleptonic
$B$ meson decay widths is written as
\begin{equation}
\Gamma=H_r(M_W,\mu)\times H(t,\mu)\times \prod_jJ_j(p_j,\mu)
\times f_B(z,\mu)\;,
\label{fat}
\end{equation}
where the structure function $f(k^+,\mu)$ has been transformed into the
distribution function $f_B(z,\mu)$ \cite{N}. The $\mu$ dependence will
disappear after performing a RG analysis, since a decay width is
scale-independent. Note that the $\mu$ dependence of the $B$ meson
distribution function for the semileptonic decays was not considered. We
shall show in the next section that such a $\mu$ dependence is not
necessary in our formalism.

\section{LOGARITHMIC SUMMATIONS}

To regularize the collinear divergences, we associate extra transverse
momentum $p_\perp$, originating from the initial $b$ quark, with the
final-state quarks, which renders them off-shell by $p_\perp$. This
transverse momentum can be regarded as the factorization scale, above which
RG evolutions are reliable, and below which dynamics is absorbed into the
$B$ meson distribution function and the jet functions. We work in the impact
parameter $b$ space, which is the Fourier conjugate variable of $p_{\perp}$.
Radiative corrections to nonleptonic decays then produce two types of large
logarithms, $\ln(M_W/t)$ and $\ln(tb)$. Especially, the double logarithms
$\ln^2({\bar p}_jb)$, ${\bar p}_j$ being the large longitudinal component of
the final-state quark momentum $p_j$, appear in the jet functions at the
kinematic end points. These logarithmic corrections should be summed to all
orders using RG equations and the resummation technique \cite{CS} in order
to improve perturbative expansions. The summation of the logarithms
$\ln(M_W/t)$ is identified as the Wilson coefficients in the effective
Hamiltonian, which describes the evolution from the characteristic scale
$M_W$ of the harder function to the characteristic scale $t$ of the hard
amplitude. The summation of the logarithms $\ln(tb)$ leads to the evolution
from $t$ to the factorization scale $1/b$.

The resummation of the double logarithms $\ln^2({\bar p}_jb)$ for a jet
with a small invariant mass is written, in the $b$ space, as \cite{ab,af}
\begin{equation}
J_j({\bar p}_j,b,\mu)=J_j(b,\mu)\exp[-2s({\bar p}_j,b)]\;.
\label{sdc}
\end{equation}
The Sudakov exponent $s$ is given by
\begin{equation}
s({\bar p}_j,b)=\int_{1/b}^{{\bar p}_j}\frac{d \mu}{\mu}
\left[\ln\left(\frac{{\bar p}_j}{\mu}
\right)A(\alpha_s(\mu))+B(\alpha_s(\mu))\right]\;,
\label{fsl}
\end{equation}
where the factors $A$ and $B$ are expanded as
\begin{eqnarray}
A(\alpha_s)&=&A^{(1)}\frac{\alpha_s}{\pi}+A^{(2)}
\left(\frac{\alpha_s}{\pi}\right)^2\;,
\nonumber \\
B(\alpha_s)&=&\frac{2}{3}\frac{\alpha_s}{\pi}\ln\left(\frac{e^{2\gamma-1}}
{2}\right)\;,
\end{eqnarray}
in order to take into account the next-to-leading-logarithm summation.
We adopt the one-loop running coupling constant,
\begin{equation}
\frac{\alpha_s(\mu)}{\pi}=\frac{1}{\beta_1\ln(\mu^2/\Lambda^2)}\;.
\end{equation}
The above coefficients $A^{(1)}$, $A^{(2)}$ and $\beta_1$ are
\begin{eqnarray}
\beta_{1}=\frac{33-2n_{f}}{12}\;,\;\;\; A^{(1)}=\frac{4}{3}\;,
\;\;\; A^{(2)}=\frac{67}{9}-\frac{\pi^{2}}{3}-\frac{10}{27}n_f
+\frac{8}{3}\beta_{1}\ln\left(\frac{e^{\gamma}}{2}\right)\; ,
\label{12}
\end{eqnarray}
where $n_f=5$ is the number of quark flavors, and $\gamma$ the Euler
constant.
The Sudakov factor $\exp(-2s)$ exhibits strong suppression at large
$b$, and approaches unity as ${\bar p}_j< 1/b$. In this region the
final-state quarks are regarded as being highly off-shell, and
absorbed into the hard amplitude. Hence, double logarithms do not exist,
{\it i.e.}, $\exp(-2s)\to 1$, and it is not necessary to introduce a jet 
function.  For the detailed derivation of Eq.~(\ref{sdc}), refer to 
\cite{af,L1}. The large scale ${\bar p}_j$ will be chosen as the sum of 
the longitudinal components of $p_j$, {\it i.e.}, ${\bar p}_j=p_j^++p_j^-$ 
in the factorization formulas presented in the next section.

The initial condition $J_j(b,\mu)$ of the resummation still contains
single logarithms $\ln(b\mu)$, which are summed by RG equations. The RG 
solution of the jet function is given by
\begin{eqnarray}
J_j(b,\mu)=J_j^{(0)}\exp\left[-\int_{1/b}^\mu\frac{d{\bar\mu}}
{\bar\mu}\gamma_j(\alpha_s({\bar\mu}))\right]\;,
\label{jet} 
\end{eqnarray}
with the anomalous dimension $\gamma_j$. It is more convenient to compute
$\gamma_j$ in the axial gauge $n\cdot A=0$, since the eikonal lines in the
direction $n$ disappear. The $n$ dependence goes into the gluon propagator
$(-i/l^2)N^{\mu\nu}(l)$, with
\begin{equation}
N^{\mu\nu}=g^{\mu\nu}-\frac{n^\mu l^\nu+n^\nu l^\mu}{n\cdot l}\;.
\label{gp}
\end{equation}
The lowest-order self-energy correction to the final-state quark is written
as
\begin{eqnarray}
\Sigma_q=-ig^2C_F\mu^{2\epsilon}\int\frac{d^{4-2\epsilon}l}
{(2\pi)^{4-2\epsilon}}\gamma_\mu
\frac{{\not p}_j-{\not l}}{(p_j-l)^2}\gamma_\nu
\frac{N^{\mu\nu}}{l^2}\frac{{\not p}_j}{p_j^2}\;,
\label{anoj}
\end{eqnarray}
which is obtained by replacing the gluon propagator $-ig^{\mu\nu}/l^2$
in Eq.~(\ref{3a}) by $-iN^{\mu\nu}/l^2$. Extracting the ultraviolet poles
from the above loop integral, we find that the $g^{\mu\nu}$ term and the
second term in Eq.~(\ref{gp}) give $-\alpha_s C_F/(4\pi)\times (1/\epsilon)$
and $\alpha_s C_F/\pi\times (1/\epsilon)$, respectively. Their sum,
$\Sigma_q=\alpha_s/\pi\times (1/\epsilon)$, leads to the quark anomalous
dimension $\gamma_q=-\alpha_s/\pi$ in the axial gauge. $\gamma_j$ is twice
of $\gamma_q$, {\it i.e.}, $\gamma_j=2\gamma_q$, because the self-energy
corrections occur before and after the final-state cut. The initial
condition $J_j^{(0)}$, with the large logarithms collected by the Sudakov
factor and by the RG evolution, can be approximated by its tree-level
expression, {\it i.e.}, the final-state cut.

The RG solution of the $B$ meson distribution function is 
\begin{eqnarray}
f_B(z,\mu)&=&f_B(z)\exp\left[-\int_{1/b}^\mu\frac{d{\bar\mu}}
{\bar\mu}\gamma_S(\alpha_s({\bar\mu}))\right]\;.
\label{sof} 
\end{eqnarray}
where the initial condition $f_B(z)$ absorbs the nonperturbative dynamics
below the scale $1/b$. The anomalous dimension $\gamma_S$ is also computed
in the axial gauge, under which the path-ordered exponential in
Eq.~(\ref{deb}) is equal to unity. Hence, we need to consider only the
self-energy correction to the $h_v$ field in Fig.~1(a). The loop integral
is written as
\begin{eqnarray}
\Sigma_v=-ig^2C_F\mu^{2\epsilon}\int\frac{d^{4-2\epsilon}l}
{(2\pi)^{4-2\epsilon}}
\frac{v_\mu}{v\cdot k}\frac{v_\nu}{v\cdot (k-l)}\frac{N^{\mu\nu}}{l^2}\;,
\label{ano}
\end{eqnarray}
where the residual momentum $k$ approaches zero. Using the
expansion
\begin{equation}
\frac{1}{v\cdot (k-l)}=-\frac{1}{v\cdot l}-\frac{v\cdot k}{(v\cdot l)^2}\;,
\end{equation}
Eq.~(\ref{ano}) reduces to
\begin{eqnarray}
\Sigma_v=ig^2C_F\mu^{2\epsilon}\int\frac{d^{4-2\epsilon}l}
{(2\pi)^{4-2\epsilon}}
\frac{v_\mu v_\nu}{(v\cdot l)^2}\frac{N^{\mu\nu}}{l^2}\;,
\label{ano1}
\end{eqnarray}
which is the same as $\Sigma^{(a)}_{\rm soft}$ in Eq.~(\ref{eik}) but with
the gluon propagator $-ig^{\mu\nu}/l^2$ replaced by $-iN^{\mu\nu}/l^2$.
Therefore, Eq.~(\ref{deb}) indeed generates the factor
$1+\Sigma^{(a)}_{\rm soft}$ in Eq.~(\ref{fas}) at first order. It is easy
to find that the contribution from the second term in Eq.~(\ref{gp})
vanishes, and that the $g^{\mu\nu}$ term gives
$\Sigma_v=\alpha_s C_F/(2\pi)\times (1/\epsilon)$.
Adding the self-energy corrections on both sides of the final-state cut,
we derive the anomalous dimension $\gamma_S=-\alpha_s C_F/\pi$ of the $B$
meson distribution function.

The RG solution of $H_r$ is 
\begin{equation}
H_r(M_W,\mu)=H_r(M_W,M_W)
\exp\left[\int_{\mu}^{M_W}\frac{d{\bar\mu}}
{\bar\mu}\gamma_{H_r}(\alpha_s({\bar\mu}))\right]\;,
\label{hrh}
\end{equation}
with $\gamma_{H_r}$ the anomalous dimension of $H_r$. The initial condition
$H_r(M_W,M_W)$ can be safely approximated by its lowest-order expression
$H^{(0)}_r=1$, since the large logarithms $\ln(M_W/\mu)$ have been
organized into the exponential, which is identified as the Wilson 
coefficient $c(\mu)$,
\begin{eqnarray}
c(\mu)\equiv \exp\left[\int_\mu^{M_W}\frac{d{\bar\mu}}{\bar\mu}
\gamma_{H_r}(\alpha_s({\bar\mu}))\right]\;.
\label{wil}
\end{eqnarray}
For the explicit expressions of the Wilson coefficients, refer to
\cite{aj}.

At last, the anomalous dimension of the hard amplitude $H$ is given by
$\gamma_H=-\gamma_{H_r}-\sum_j\gamma_j-\gamma_S$ because of the scale
invariance of a decay width in the full theory. Applying the RG analysis,
we obtain
\begin{eqnarray}
H(t,\mu)=H(t,t)\exp\Bigg\{-\int_{\mu}^t\frac{d{\bar\mu}}{\bar\mu}
\left[\gamma_{H_r}(\alpha_s({\bar\mu}))+\sum_j\gamma_j(\alpha_s({\bar\mu}))+
\gamma_S(\alpha_s({\bar\mu}))\right]\Bigg\}\;.
\label{h}
\end{eqnarray}
The scale $t$ will be chosen as the maximal relevant scales,
\begin{equation}
t = \max \left( {\bar p}_j, \frac{1}{b} \right)\;.
\label{tsc}
\end{equation}
Since the large $b$ region is Sudakov suppressed by $\exp(-2s)$ as stated
before, that is, $t$ remains as a hard scale, and the large logarithms
$\ln(t/\mu)$ in $H$ have been grouped into the exponential, the initial 
condition $H(t,t)$ is calculable in perturbation theory. Hence, the Sudakov 
factor, though important only in the end-point region, improves the 
applicability of PQCD to inclusive nonleptonic heavy hadron decays.

Substituting Eqs.~(\ref{sdc}), (\ref{jet}), (\ref{sof}), (\ref{hrh}) and
(\ref{h}) into Eq.~(\ref{fat}), we derive the RG improved factorization
formula for the nonleptonic $B$ meson decay widths,
\begin{equation}
\Gamma=c(t)H(t,t)f_B(z)\exp\Bigg\{-\sum_j\left[
2s({\bar p}_j,b)+\int_{1/b}^t\frac{d{\bar\mu}}{\bar\mu}
\gamma_j(\alpha_s({\bar\mu}))\right]
-\int_{1/b}^t\frac{d{\bar\mu}}{\bar\mu}
\gamma_S(\alpha_s({\bar\mu}))\Bigg\}\;,
\label{main}
\end{equation}
where the cancellation of the $\mu$ dependences among the the convolution
factors is explicit, and the two-stage evolutions from $1/b$ to $t$ and
from $t$ to $M_W$ have been established. Note that Eq.~(\ref{main}) in fact
denotes the integrand appearing in the factorization formula, and that the
variables $t$ and $b$ will be integrated out. The initial conditions
$J_j^{(0)}$ in Eq.~(\ref{jet}) have been incorporated into the evaluation of
$H(t,t)$. We emphasize that the Wilson coefficient appears as a convolution
factor in the three-scale factorization theorem, instead of a constant (once
its argument $\mu$ is set to a common value for all decay modes) in the
HQET approach. Since the perturbative factors depend on the
hadron kinematics, they vary in different decay modes of a hadron and in
different hadrons. It will be shown that these perturbative effects play an
essential role in the explanation of the $b$-hadron lifetimes.

In the semileptonic case there are not the Wilson coefficient $c(t)$
and the Sudakov factor $\exp(-2s)$ because of the absence of the double
logarithms. Therefore, the large scales ${\bar p}_j$ do not exist, and
$t$ is equal to $1/b$, for which the RG evolution factors governed by
the anomalous dimensions $\gamma_j$ and $\gamma_S$ disappear. This is the 
reason we  did not perform the RG analysis for the semileptonic decays in
Sec.~II.

\section{FACTORIZATION FORMULAS AND NUMERICAL ANALYSIS}

After developing the factorization theorems, we present the 
factorization formulas for the $B$ meson semileptonic and nonleptonic
decay widths. Define the momentum of the light degrees of freedom as
$p=(p^+,0,{\bf p}_\perp)$. The $B$ meson is at rest with the
mometum $P_B=M_B/\sqrt{2}(1,1,{\bf 0})$. The $b$ quark momentum is then
written as $p_b=P_B-p=(zM_B/\sqrt{2},M_B/\sqrt{2},-{\bf p}_\perp)$, where 
the momentum fraction $z\equiv p_b^+/P_B^+=1-\sqrt{2}p^+/M_B$ is the same 
as that defined in Sec. III. The lepton and neutrino momenta involved in the
semileptonic decays $B(P_{B}) \rightarrow X_{c}+l(p_l)+\bar{\nu}(p_{\nu})$
are expressed, in terms of light-cone coordinates, as
\begin{equation}
     p_l=(p^+_l,p_l^-,0_{\perp})\;,\mbox{\ \ }
     p_{\nu}=(p_{\nu}^+,p_{\nu}^-,{\bf p}_{\nu \perp})\;,
\end{equation}
where the minus component $p_l^-$ vanishes for a massless lepton.
For convenience, we adopt the scaling variables,
\begin{equation}
x=\frac{2E_l}{M_B}\;,\mbox{\ \ } y=\frac{q^2}{M_B^2}\;,\mbox{\ \ }
y_0=\frac{2q_0}{M_B}\;,
\label{dl}
\end{equation}
with the kinematic ranges,
\begin{eqnarray}
2\sqrt{\alpha}\leq&x& \leq 1+\alpha-\beta\;,
\nonumber\\
\alpha\leq &y& \leq \alpha+(1+\alpha-\beta-x)
\frac{x+\sqrt{x^2-4\alpha}}{2-x-\sqrt{x^2-4\alpha}}\;,
\nonumber\\
x+\frac{2(y-\alpha)}{x+\sqrt{x^2-4\alpha}}\leq &y_0&\leq 1+y-\beta\;,
\label{kcs}
\end{eqnarray}
where $E_l$ is the lepton energy and $q \equiv p_l+p_{\nu}$ 
the lepton pair momentum. The constants $\alpha$ and $\beta$ are
\begin{equation}
\alpha \equiv \frac{M^2_l}{M^2_B}\;,\;\;\;\;
\beta \equiv \frac{M^2_D}{M^2_B}\;,
\end{equation}
$M_l$ and $M_D$ being the lepton mass and the $D$ meson mass, respectively.
$M_D$ arises as the minimal invariant mass of the decay product $X_c$.
For the derivation of Eq.~(\ref{kcs}), refer to the Appendix.

The factorization formula for the semileptonic
decay width is given, in the $b$ space, by \cite{ai}
\begin{equation}
\frac{\Gamma_{\rm SL}}{\Gamma_0}=\frac{M_B^2}{2\pi}\int dxdydy_0
\int^1_{z_{\rm min}}{dz}
\int_0^\infty db b f_B(z){\tilde J}_c(x,y,y_0,z,b)H(x,y,y_0,z)\;,
\label{asb}
\end{equation}
with $\Gamma_0 \equiv (G_F^2/16\pi^3)|V_{cb}|^2M^5_B$. The momentum
fraction $z$ approaches 1 as the $b$ quark carries the whole $B$ meson
momentum in the plus direction. The minimum of $z$, determined by the
condition $p_c^2 > M_c^2$, is
\begin{equation}
z_{\rm min} =
\frac{\displaystyle\frac{y_0}{2}-y+\frac{M_c^2}{M_B^2}-
\frac{x}{\sqrt{x^2-4\alpha}} \left[-\frac{y_0}{2}+\frac{y}{x}+
\frac{\alpha}{x}\right] }
{\displaystyle 1-\frac{y_0}{2}-\frac{x}{\sqrt{x^2-4\alpha}}
\left[-\frac{y_0}{2}+\frac{y}{x}+\frac{\alpha}{x} \right] }\;.
\end{equation}
The function ${\tilde J}_c$ denotes the Fourier transformation of the
final-state cut on the $c$ quark line, which is in fact part of the hard
amplitude. $J_c$ and the lowet-order $H$ in momentum space are
\cite{ai}
\begin{eqnarray}
J_c&=&\delta(p_c^2-M_c^2)\;,
\nonumber\\
&=& \delta\left( M_B^2\left\{z-(1+z)\frac{y_0}{2}+y+
\frac{x(1-z)}{\sqrt{x^2-4\alpha}}
\left[-\frac{y_0}{2}+\frac{y}{x}+\frac{\alpha}{x}\right]
\right \}-M_c^2-{\bf p}^2_{\perp}\right),
\nonumber \\
& &\label{jc}\\
H&=&(p_b\cdot p_\nu)(p_l\cdot p_c)\;,
\nonumber\\
&=&\left((y_0-x)\left\{1-\frac{(1-z)}{2}
\left(1-\frac{x}{\sqrt{x^2-4\alpha}}\right)\right\}
-\frac{(1-z)}{\sqrt{x^2-4\alpha}}
(y-\alpha)\right)\nonumber\\
&&\times \left(\frac{x}{2}\left\{1+z+(1-z)
\frac{\sqrt{x^2-4\alpha}}{x}\right\}
-y-\alpha\right).
\label{hi}
\end{eqnarray}
The universal $B$ meson distribution function $f_B$, determined from the
$B \rightarrow X_s \gamma$ decay, has been given in Eq.~(\ref{bdf}),
which minimizes the model dependence of our predictions.

For the nonleptonic decays, we consider the modes $b\to c{\bar c}s$
and $b\to c{\bar u}d$. Ignoring the penguin operators, the effective 
Hamiltonian for the $b\to c{\bar c}s$ decay is
\begin{equation}
H_{\rm eff} = \frac{4G_{F}}{\sqrt{2}} V_{cb} V^{\ast}_{cs}
          [ \, c_{1}(\mu) O_{1}(\mu) +
            c_{2}(\mu) O_{2}(\mu) \, ]\;,
\label{eff1}
\end{equation}
with $G_F$ the Fermi coupling constant, $V$'s the 
Cabibbo-Kabayashi-Maskawa (CKM) matrix elements, the four-quark
operators $O_{1}=(\bar{s}b)(\bar{c}c)$ and $O_{2}=(\bar{c} b)(\bar{s}c)$,
and the initial conditions $c_1(M_W)=1$ and $c_2(M_W)=0$.
For the $b\to c{\bar u}d$ decay, $V_{cs}$ and the ${\bar c}$ and $s$ quark
fields are replaced by $V_{ud}$ and the ${\bar u}$ and $d$ quark fileds,
respectively. It is simpler to work with the operators
$O_{\pm}=\frac{1}{2} (O_{2} \pm O_{1})$ and their corresponding
coefficients $c_{\pm}(\mu) = c_{2}(\mu) \pm c_{1}(\mu)$, since
they are multiplicatively renormalized. 
In the leading logarithmic approximation $c_{\pm}$ are given by \cite{aj}
\begin{equation}
c_{\pm}(\mu) = \left[ \frac{\alpha_{s}(M_{W})}{\alpha_{s}(\mu)} \right]
^{\frac{-6\gamma_{\pm}}{33-2n_{f}}} \;,
\end{equation}
with the constants $2\gamma_+=-\gamma_-=-2$ and $n_f=5$.

To simplify the analysis, we route the transverse momentum $p_\perp$ of the
$b$ quark through the outgoing $c$ quark as in the semileptonic case, such
that the $c$ quark momentum is the same as before. For kinematics, we make
the correspondence with the ${\bar c}$ (${\bar u}$) quark carrying the
momentum of the massive (light) lepton $\tau$ ($e$ and $\mu$) and with the
$s$ and $d$ quarks carrying the momentum of ${\bar\nu}$. The scaling
variables are then defined exactly by Eq.~(\ref{dl}). The factorization
formula for the nonleptonic decay widths is written, according to the
three-scale factorization theorem in Sec. IV, as \cite{ai}
\begin{eqnarray}
\frac{\Gamma_{\rm NL}}{\Gamma_0}
&=&\frac{M_B^2}{2\pi}\int dxdydy_0\int_{z_{\rm min}}^1 dz
\int_0^\infty bdb \left[ \frac{N_c\!+\!1}{2}
c_{+}^2(t) + \frac{N_c\!-\!1}{2} c_{-}^2(t) \right]
\nonumber \\
& & \times f_B(z){\tilde J}_c(x,y,y_0,z,b)H(x,y,y_0,z)
S({\bar p}_j,t,b)\;.
\label{non}
\end{eqnarray}
The $B$ meson distribution function $f_B$, being universal, is the same as
that for the semileptonic decays. The factor $S$ is the result of the
Sudakov resummation and the RG evolutions, given by
\begin{equation}
S({\bar p}_j,t,b)=\exp\Bigg\{-\sum_j\left[
2s({\bar p}_j,b)+\int_{1/b}^t\frac{d{\bar\mu}}{\bar\mu}
\gamma_j(\alpha_s({\bar\mu}))\right]
-\int_{1/b}^t\frac{d{\bar\mu}}{\bar\mu}
\gamma_S(\alpha_s({\bar\mu}))\Bigg\}\;,
\end{equation}
with $j=s$ for the $b \rightarrow c\bar{c}s$ mode and $j={\bar u}$, $d$ for
the $b \rightarrow c\bar{u}d$ mode. The explicit expressions of
${\bar p}_j$ are
\begin{eqnarray}
{\bar p}_s={\bar p}_d=\frac{M_B}{\sqrt{2}}(y_0-x)\;,
\;\;\;\;
{\bar p}_u=\frac{xM_B}{\sqrt{2}}\;.
\end{eqnarray}
The anomalous dimensions $\gamma_j=-2\alpha_s/\pi$ and
$\gamma_S=-C_F\alpha_s/\pi$ have been computed in Sec. V. It has been
found that the single-logarithm evolutions governed by $\gamma_j$ and 
$\gamma_S$ enhance the nonleptonic branching
ratios and thus lower the semileptonic branching ratios \cite{ai},
consistent with the observation in \cite{bagan}. 

The above formalism can be generalized to the $B_s$ meson and $\Lambda_b$
baryon decays straightforwardly, for which the functional forms of the hard
amplitudes $H$, the final-state cut $J_c$, the Wilson coefficients 
$c_\pm$, and the Sudakov evolution factor $S$ remain the same, since they
are evaluated at the quark level. The only differences arise from the
replacement of the $B_d$ meson mass $M_B$ by the $B_s$ meson mass $M_{B_s}$
and by the $\Lambda_b$ baryon mass $M_{\Lambda_b}$, and from the heavy
hadron distribution functions given in Eqs.~(\ref{bds}) and (\ref{bdb}).
We then proceed with the numerical analysis of the factorization formulas
in Eqs.~(\ref{asb}) and (\ref{non}) for the semileptonic and nonleptonic 
$b$-hadron decays, respectively, choosing the CKM matrix elements 
$|V_{cs}|=|V_{ud}|=1.0$ and $|V_{cb}|=0.044$, and
the masses $M_c=1.6$ GeV, $M_D=1.869$ GeV and $M_\tau=1.7771$ GeV.
Our predictions are not sensitive to the change of
$M_c$: the difference of the results for $M_c=1.5$ GeV from those for
$M_c=1.6$ GeV is less than 5\%. We obtain the lifetimes
$\tau(B_d)=1.56$ ps, $\tau(B_s)=1.46$ ps and $\tau(\Lambda_b)=1.22$ ps,
or the ratios $\tau(B_s)/\tau(B_d)=0.94$ and
$\tau(\Lambda_b)$/$\tau(B_d)=0.78$.

The experimental data of the $b$-hadron lifetimes are summarized below.
The lifetimes of the $B_d$ and $B_s$ mesons and
of the $\Lambda_b$ baryon are $\tau(B_d)=(1.55\pm 0.03)$ ps,
$\tau(B_s)=(1.47\pm 0.06)$ ps and
$\tau(\Lambda_b)=(1.23\pm 0.05)$ ps, respectively, from the
CERN $e^+e^-$ collider LEP measurements \cite{ad}. Recent CDF results yield
$\tau(\Lambda_b)=(1.32\pm 0.16)$ ps \cite{al}. The lifetime
ratios are then $\tau(B_s)/\tau(B_d)=0.95\pm 0.06$ and
$\tau(\Lambda_b)$/$\tau(B_d)=0.79\pm 0.06$ from LEP, and
$\tau(\Lambda_b)$/$\tau(B_d)=0.85\pm 0.11$ from CDF. Obviously, our
predictions are almost the same as the central values of the LEP data.
Employing the $B_u$ meson distribution function in Eq.~(\ref{bdu}), we
obtain $\tau(B_u)=1.62$ ps or the ratio $\tau(B_u)/\tau(B_d)=1.04$, which is
in agreement with the experimental data $\tau(B_u)=1.65\pm 0.04$ ps in
\cite{C}.

To test the sensitivity of our predictions to the variation of
$\lambda_{1B_s}$ and $\lambda_{1\Lambda_b}$, we adopt the HQET parameters
$\lambda_1^{\rm meson}\sim\lambda_1^{\rm baryon}=-0.4$ GeV$^2$
from QCD sum rules for $\lambda_{1B_s}$ and $\lambda_{1\Lambda_b}$
(with the correspondiong ${\bar\Lambda}_{B_s}=0.77$ GeV and 
${\bar\Lambda}_{\Lambda_b}=1.03$ GeV). The lifetimes $\tau(B_s)=1.49$ ps
and $\tau(\Lambda_b)=1.26$, or the ratios $\tau(B_s)/\tau(B_d)=0.96$ and 
$\tau(\Lambda_b)$/$\tau(B_d)=0.81$, are derived, which are still much 
smaller than those from the HQET approach. This test indicates that our 
results are insensitive to the variation of the distribution functions 
(less than 4\% under 40\% variation of the HQET parameters), and that the
explanation of the experimental data is mainly due to the PQCD effects. 
To confirm this observation, we set all the $b$-hadron masses in the phase 
space factors $\Gamma_0$ and in the perturbative factors to the $b$ quark 
mass $M_b$ in Eqs.~(\ref{asb}) and (\ref{non}). That is, the difference 
among the factorization formulas for $b$-hadron decays resides only in the 
distribution functions. The lifetime ratios $\tau(B_s)/\tau(B_d)=1.01$ and 
$\tau(\Lambda_b)/\tau(B_d)=1.09$, which are even greater than unity, are 
obtained. Therefore, the perturbative contributions are indeed responsible
for the low $b$-hadron lifetime ratios.

The above results are aummarized in Table I. In Table I we also present the
branching ratio of each mode in the $B_d$ and $B_s$ meson decays and in the
$\Lambda_b$ baryon decays in terms of the quantities
$r_{\tau\nu}=B(b\to c\tau{\bar \nu})/B(b\to cl{\bar \nu})$,
$r_{ud}=B(b\to c{\bar u}d)/B(b\to cl{\bar \nu})$, 
$r_{cs}=BR(b\to c{\bar c}s)/B(b\to cl{\bar \nu})$,
the semileptonic branching ratio $B_{\rm SL}$ and the average
charm yield $\langle n_c\rangle$ per $b$-hadron decay. It is observed that
our predictions of $B_{\rm SL}=10.16\%$ and $\langle n_c\rangle=1.17$ for
the $B$ meson are consistent with the experimental data:
$B_{\rm SL}=(10.19\pm 0.37)$ and $\langle n_c\rangle=(1.12\pm 0.05)$
from the CLEO group\cite{ak}, and $B_{\rm SL}=(11.12\pm 0.20)$ and
$\langle n_c\rangle=(1.20\pm 0.07)$ from the LEP measurements \cite{LEP}.
It is also observed that $B_{\rm SL}$, $\langle n_c\rangle$ and all $r$'s
except $r_{ud}$ increase a bit with the $b$-hadron masses. It is interesting
to examine these tendencies in future experiments.

\section{CONCLUSION}

In this paper we have developed the PQCD factorization theorems
for inclusive $b$-hadron decays by carefully analyzing the infrared
divergences in radiative corrections to $b$ quark decays. Radiative
corrections characterized by the hadronic scale are absorbed into the heavy
hadron distribution function or into the jet functions. Above the hadronic
scale, perturbative contributions characterized by the $b$-hadron mass and
the $W$ boson mass are absorbed into the hard $b$ quark decay amplitude and
the harder function, respectively. Various large logarithmic corrections
have been summed to all orders using the resummation technique and the RG
equations, leading to the evolution factors among the three characteristic
scales, such that the factorization formulas are $\mu$-independent. We have
shown that the soft cancellation between virtual and real corrections
demands the introduction of the heavy hadron distribution function, and thus
the $b$-hadron kinematics into our formalism. Using the PQCD factorization
theorems, we have been able to explain the lifetimes of the $b$-hadrons
$B_d$, $B_u$, $B_s$ and $\Lambda_b$. It has been confirmed that our
predictions are insensitive to the variation of the distribution functions
as expected from the quark-hadron duality, and that the low $b$-hadron
lifetime ratios are mainly attributed to the perturbative contributions.

\vskip 0.5cm

I thank Profs. H.Y. Cheng and H.L. Yu for useful discussions. This work
was supported by the National Science Council of R.O.C. under
Grant No. NSC 88-2112-M-006-013.
\vskip 1.0cm

\centerline{\bf APPENDIX}
\vskip 0.5cm

In this appendix we derive the phase space constraints for the decay
$b\to cl{\bar\nu}$. The constraints for the decay $b\to c{\bar c}s$ are
the same with the ${\bar c}$ and $s$ quarks corresponding to the $l$ and
${\bar\nu}$ leptons, respectively. The constraints for the decay
$b\to c{\bar u}d$ is the massless lepton case of the results presented
below.

We derive the constraint of $x=2E_l/M_H$ from the energy and momentum
conservations associated with the decay $H\to X_cl{\bar\nu}$,
\begin{eqnarray}
M_H&=&E_X+E_l+E_\nu\;,
\nonumber\\
0&=&{\bf p}_X+{\bf p}_l+{\bf p}_\nu\;.
\label{con}
\end{eqnarray}
Because of $E_l^2=|{\bf p}_l|^2+M_l^2$, the minimum of $E_l$ occurs at the
minimum of $|{\bf p}_l|$, $|{\bf p}_l|=0$, and thus the minimum of $x$ is
given by
\begin{equation}
x_{\min}=\frac{2M_l}{M_H}\;.
\label{xmin}
\end{equation}
To obtain the maximum of $x$, we find the maximum of $|{\bf p}_l|$ under
Eq.~(\ref{con}),
\begin{equation}
M_H=\sqrt{({\bf p}_l+{\bf p}_\nu)^2+M_X^2}+
\sqrt{|{\bf p}_l|^2+M_l^2}+|{\bf p}_\nu|^2\;,
\end{equation}
into which ${\bf p}_X=-({\bf p}_\tau+{\bf p}_\nu)$ has been inserted. It is
easy to observe that the maximum of $|{\bf p}_l|$ corresponds to
${\bf p}_\nu=0$, leading to
\begin{equation}
|{\bf p}_{l\max}|^2=\frac{(M_H^2+M_l^2-M_X^2)^2-4M_H^2M_l^2}{4M_H^2}\;,
\end{equation}
and
\begin{equation}
x_{\max}=1+\frac{M_l^2}{M_H^2}-\frac{M_X^2}{M_H^2}\;.
\label{xmam}
\end{equation}
Combining Eqs.~(\ref{xmin}) and (\ref{xmam}), we obtain
\begin{equation}
\frac{2M_l}{M_H}\le x\le
1+\frac{M_l^2}{M_H^2}-\frac{M_D^2}{M_H^2}\;,
\end{equation}
where $M_X$ has been replaced by the $D$ meson mass $M_D$, the mass of the
lightest charmed hadron.

We then derive the constraint of the kinematic variable $y=q^2/M_H^2$ with
\begin{equation}
q^2=(p_l+p_\nu)^2=M_l^2+2(E_l-|{\bf p}_l|\cos\theta)E_\nu\;,
\label{q2}
\end{equation}
where $\theta$ is the angle between the vectors ${\bf p}_l$ and
${\bf p}_\nu$. The minimum of $q^2$ occurs, as $\cos\theta=1$ and
$E_\nu$ takes its minimal value under the constraint from Eq.~(\ref{con}),
\begin{equation}
M_H-E_l=\sqrt{|{\bf p}_l|^2+E_\nu^2+2|{\bf p}_l|
E_\nu\cos\theta+M_X^2}+E_\nu\;.
\label{cons2}
\end{equation}
Obviously, the minimum of $E_\nu$ is zero, which can be achieved by
increasing $M_X$. We then have
\begin{equation}
(q^2)_{\min}=M_l^2\;.
\label{ymin}
\end{equation}
On the other hand, the maximum of $q^2$ occurs, as $\cos\theta=-1$ and
$E_\nu$ takes its maximal value under Eq.~(\ref{cons2}), which corresponds
to the minimum of $M_X$. Setting $M_X=M_D$, we obtain
\begin{equation}
E_{\nu\max}=\frac{M_H^2+M_l^2-M_D^2-2M_HE_l}
{2(M_H-E_l-\sqrt{E_l^2-M_l^2})}\;,
\end{equation}
and
\begin{equation}
(q^2)_{\max}=M_l^2+(M_H^2+M_l^2-M_D^2-2M_HE_l)
\frac{E_l+\sqrt{E_l^2-M_l^2}}
{M_H-E_l-\sqrt{E_l^2-M_l^2}}\;.
\label{ymax}
\end{equation}
Combining Eqs.~(\ref{ymin}) and (\ref{ymax}), we have
\begin{equation}
\frac{M_l^2}{M_H^2}\le y\le
\frac{M_l^2}{M_H^2}
\left(1+\frac{M_l^2}{M_H^2}-\frac{M_D^2}{M_H^2}-x\right)
\left(\frac{2}{2-x-\sqrt{x^2-4M_l^2/M_H^2}}-1\right)\;.
\end{equation}

At last, we derive the range of the variable $y_0=2q_0/M_H$ with
\begin{equation}
y_0=E_l+E_\nu\;,
\end{equation}
or equivalently, the range of $E_\nu$ under the constraints of
Eqs.~(\ref{q2}) and (\ref{cons2}). Since $E_\nu$ decreases with
$\cos\theta$ in order that $q^2$ maintains constant, the minimum of $E_\nu$
occurs at $\cos\theta=-1$. Equation (\ref{q2}) then gives
\begin{equation}
E_{\nu\min}=\frac{q^2-M_l^2}{2(E_l+\sqrt{E_l^2-M_l^2})}\;,
\label{evm}
\end{equation}
and
\begin{equation}
y_{0\min}=x+\frac{2(y-M_l^2/M_H^2)}{x+\sqrt{x^2-4M_l^2/M_H^2}}\;.
\label{ymi}
\end{equation}
It is easy to observet that $E_\nu$ increases as $M_X$ decreases.
Setting $M_X=M_D$, and substituting the expression of $\cos\theta$ from
Eq.~(\ref{q2}) into (\ref{cons2}), we obtain
\begin{equation}
E_{\nu\max}=\frac{M_H^2-2M_HE_l-M_D^2+q^2}{2M_H}\;,
\label{evmx}
\end{equation}
and
\begin{equation}
y_{0\max}=1+y-\frac{M_D^2}{M_H^2}\;.
\label{yma}
\end{equation}
Combining Eqs.~(\ref{ymi}) and (\ref{yma}), we derive
\begin{equation}
x+\frac{2(y-M_l^2/M_H^2)}{x+\sqrt{x^2-4M_l^2/M_H^2}}
\le y_0\le 1+y-\frac{M_D^2}{M_H^2}\;.
\end{equation}

\newpage

Table I. The $b$-hadron lifetimes $\tau(B_d)$, $\tau(B_u)$, $\tau(B_s)$,
and $\tau(\Lambda_b)$, and their ratios to $\tau(B_d)$. Predictions of
$r_{\tau\nu}$, $r_{ud}$, $r_{cs}$, $B_{\rm SL}$, and $\langle n_c\rangle$
for the $b$-hadron decays are also listed.

\vskip 0.5cm

\begin{center}
\begin{tabular}{lccccccc}
\hline
             & $r_{\tau \nu }$ & $r_{ud}$ & $r_{cs}$ &
$B_{\rm SL}$ & $\langle n_{c}\rangle $ & $\tau$ (ps) & $\tau/\tau(B_d)$ \\
\hline
$\lambda_{1}=-0.71$ GeV$^2$
& 0.224 & 5.98 & 1.64 & 10.16 & 1.17 &1.56 &1.0 \\
\hline
$\lambda_{1B_u}=-0.81$ GeV$^2$
& 0.222 & 6.11 & 1.54 & 10.14 & 1.16 &1.62 &1.04 \\
\hline
$\lambda_{1B_s}=-0.4$ GeV$^2$ & 0.231 & 5.25 & 1.44 & 11.21 &
1.16 &1.49 & 0.96\\
\hline
$\lambda_{1B_s}=-0.71$ GeV$^2$ & 0.237 & 5.30 & 1.69 & 10.85 &
1.18 & 1.46 &0.94\\
\hline
$\lambda_{1\Lambda_b}=-0.4$ GeV$^2$ & 0.254 & 5.14 & 1.75 & 10.94 &
1.19 &1.26 & 0.81\\
\hline
$\lambda_{1\Lambda_b}=-0.71$ GeV$^2$ & 0.261 & 5.26 & 1.68 & 10.87 &
1.18 & 1.22 &0.78\\
\hline
\end{tabular}
\end{center}

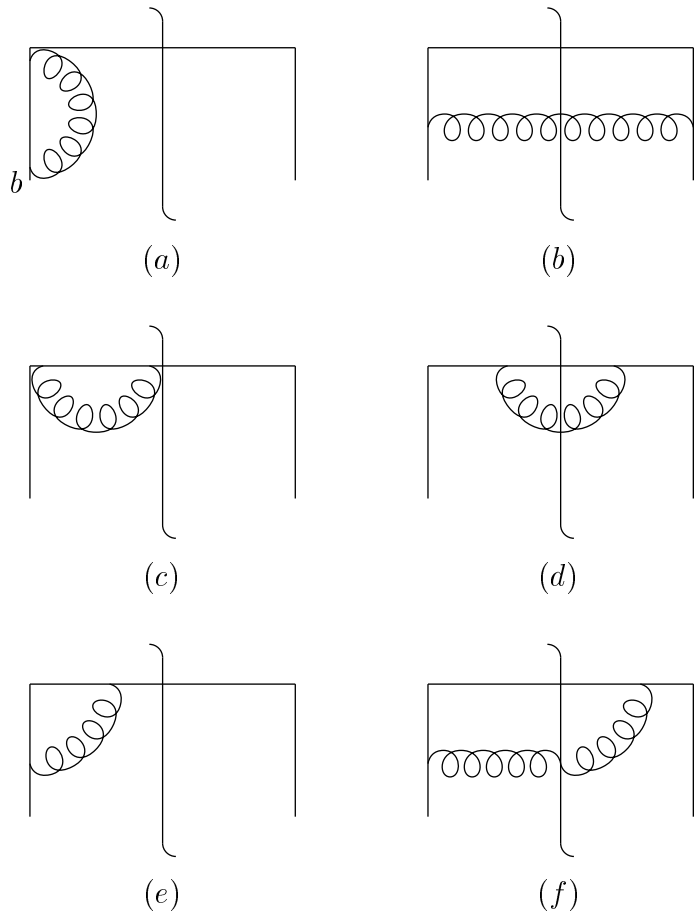
\end{document}